\documentclass[aps,prl,twocolumn,superscriptaddress]{revtex4}
\usepackage{amssymb}
\usepackage{graphicx}
\usepackage{amsmath}
\usepackage{verbatim}
\usepackage{subfigure}
\usepackage{amsmath}
    \usepackage{amssymb}
    \usepackage{amsthm}
    \usepackage{amsfonts}
    \usepackage{braket}

\begin{document}
	
	\title{Interaction effects on $\mathcal{PT}$-symmetry breaking transition in atomic gases}
	\author{Ziheng Zhou}
	\affiliation{Laboratory of Quantum Engineering and Quantum Metrology, School of Physics and Astronomy, Sun Yat-Sen University (Zhuhai Campus), Zhuhai 519082, China}
	
	\author{Zhenhua Yu}
	\email[]{huazhenyu2000@gmail.com}
	\affiliation{Laboratory of Quantum Engineering and Quantum Metrology, School of Physics and Astronomy, Sun Yat-Sen University (Zhuhai Campus), Zhuhai 519082, China}
	\affiliation{State Key Laboratory of Optoelectronic Materials and Technologies,
Sun Yat-Sen University (Guangzhou Campus), Guangzhou 510275, China}
	\date{\today }
	
	\begin{abstract}
	Non-Hermitian systems having parity-time ($\mathcal {PT}$) symmetry can undergo a transition, spontaneously breaking the symmetry. Ultracold atomic gases provide an ideal platform to study interaction effects on the transition. We consider a model system of $N$ bosons of two components confined in a tight trap. Radio frequency and laser fields are coupled to the bosons such that the single particle Non-Hermitian Hamiltonian $h_{\mathcal PT}=-i \Gamma\sigma_z+J\sigma_x$, which has $\mathcal {PT}$-symmetry, can be simulated in a \emph{passive} way. We show that when interatomic interactions are tuned to maintain the symmetry, the $\mathcal {PT}$-symmetry breaking transition is affected only by the SU(2) variant part of the interactions parameterized by $\delta g$. We find that the transition point $\Gamma_{\rm tr}$ decreases as $|\delta g|$ or $N$ increases; in the large $|\delta g|$ limit, $\Gamma_{\rm tr}$ scales as $\sim|\delta g|^{-(N-1)}$. We also give signatures of the 
$\mathcal {PT}$-symmetric and the symmetry breaking phases for the interacting bosons in experiment.	
	\end{abstract}
	
	\maketitle

Study of non-Hermitian systems is constantly enriching our knowledge derived from Hermitian ones \cite{Bender, 
Muss,Moi,Guo,Kip,Ueda,Segev,Khaj,Fu}. Of particular interest are a class of non-Hermitian systems having the parity-time ($\mathcal{PT}$) symmetry \cite{Bender1999, Moibook, Benderbook}. A representative model of the class is a two-level system whose Hamiltonian is of the form $h_{\mathcal PT}=-i \Gamma\sigma_z+J\sigma_x$; under the combined transformation of complex conjugate and the swap $\ket\uparrow\leftrightarrow\ket\downarrow$, $h_{\mathcal PT}$ is invariant \cite{Swanson}. Parameter tuning across the critical point $\Gamma_{\rm tr}=J$ gives rise to the transition of the two-level system from the $\mathcal {PT}$-symmetric phase to the symmetry breaking phase where exponentially growing or decaying modes set in. $\mathcal{PT}$-symmetry breaking transition has been widely investigated in electromagnetic \cite{Guo,Kip, Kottos, Peschel, Schafer, Yang}, and mechanical systems \cite{Bender2013}. The transition is the cornerstone of exceptional properties regarding light propagation \cite{Lin2011, Scherer2011, Scherer2013}, lasing \cite{Yang2014, Khaj2014, Zhang2014} and topological energy transfer \cite{Rotter2016, Harris2016}.

Recently $\mathcal{PT}$-symmetry breaking transition was successfully demonstrated in a gas of two component noninteracting $^6$Li atoms in a \emph{passive} way \cite{Luo}; in the experiment, the application of a radio-frequency field and a laser inducing loss in one component of the atoms leads to, apart from kinetic energy, the single particle Hamiltonian $h=-i\Gamma+h_{\mathcal PT}$, as the term $-i\Gamma$ gives rise to an overall decay. This approach circumvents the difficulty of realizing an atom gain in quantum simulation of $\mathcal{PT}$-symmetric non-Hermitian Hamiltonians in atomic gases \cite{Moi}. On the other hand, Feshbach resonance enables unprecedented control of interactions in ultracold atomic gases \cite{Chin}, and deterministic preparation is achievable for a sample of variable $N$ atoms \cite{Bloch2010, Greiner2010, Bloch2011, Jochim}. These capabilities make ultracold atoms an ideal platform to probe interaction effects on $\mathcal{PT}$-symmetry breaking transition \cite{Konotop2016}.

In this work, we consider $N$ interacting two component bosons confined in a tight harmonic trap such that their spatial wave-function is frozen to be the ground harmonic state. The bosons are subject to the radio frequency field and the laser as in Ref. \cite{Luo}. Feshbach resonance is used to tune the interaction Hamiltonian of the $N$ bosons to maintain the $\mathcal PT$-symmetry. We find that in this interacting system, $\mathcal{PT}$-symmetry breaking transition depends on only the SU(2) variant part of the interactions parameterized by $\delta g$. The transition point $\Gamma_{\rm tr}$ decreases as $|\delta g|$ or $N$ increases. In the large $|\delta g|$ limit, $\Gamma_{\rm tr}$ is suppressed as $\sim|\delta g|^{-(N-1)}$. Finally we show how the modification on the transition by the interactions can be detected experimentally.

\begin{figure}[t]
			\includegraphics[width=3 in]{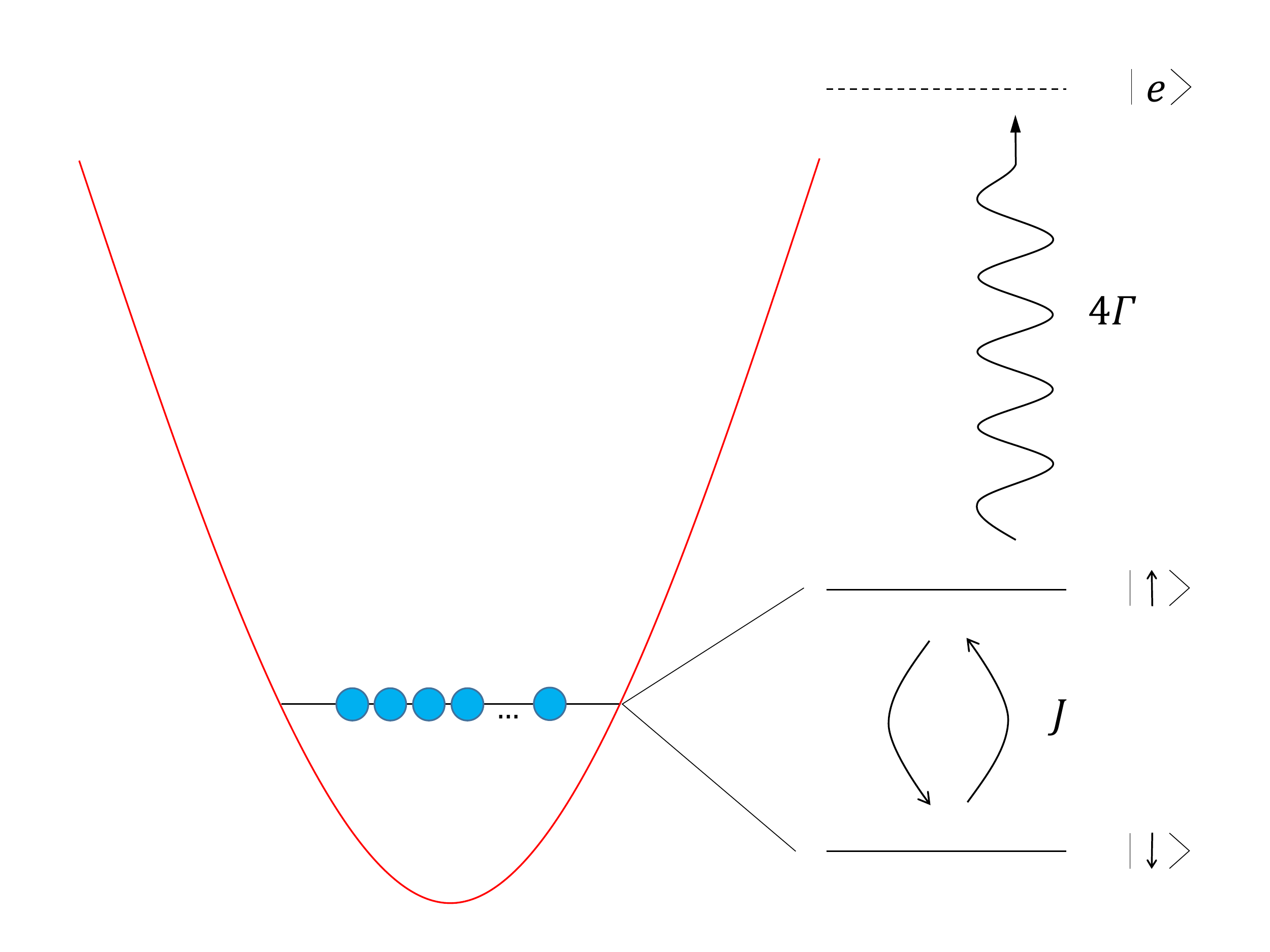}
			\caption{Schematic of the system. Bosons are spatially confined in the single particle ground state of a tight harmonic potential. A resonant radio-frequency field couples two internal states of each bosons, $\ket\uparrow$ and $\ket\downarrow$, with Rabi frequency $J$. An additional laser couples the internal state $\ket\uparrow$ to an another excited state $\ket e$ and results in a number loss in state $\ket\uparrow$ of rate $4\Gamma$.}
			\label{setup}
		\end{figure}

Figure (\ref{setup}) gives a schematic of the system that we consider. Bosonic atoms with two internal states denoted by $\ket\uparrow$ and $\ket\downarrow$ are confined in a harmonic trap potential $V(\mathbf r)=\frac12 m\omega_0^2r^2$, where $m$ is the atomic mass. For simplicity, we assume the confinement being so tight, i.e., $\omega_0$ is much larger than any other energy scales to be considered, that the spatial wave-function of the bosons is frozen to be the single particle ground state $\phi_0(\mathbf r)$ of the harmonic trap. A radio-frequency field of frequency equal to the internal energy difference $E_\uparrow-E_\downarrow$ is used to flip the atoms between the two internal states ${\ket\uparrow}$ and $\ket\downarrow$ with Rabi frequency $J$. An additional laser is coupled to the atoms in state $\ket\uparrow$ and results in a loss rate $4\Gamma$ of the atom number in the state. We take $\hbar=1$ throughout.

In the absence of interatomic interactions, the Hamiltonian for each bosons spanned by $\ket\uparrow$ and $\ket\downarrow$ is non-Hermitian and is given by $h=-i\Gamma+h_{\mathcal PT}$, apart from the ground harmonic state energy \cite{Luo}. 
The interatomic interaction Hamiltonian of the bosons is given by
\begin{align}
H_{int}=\sum_{j\neq k}&\left[\frac{g_{\uparrow\uparrow}}2\frac{\sigma^{(j)}_z+1}2\frac{\sigma^{(k)}_z+1}2+\frac{g_{\downarrow\downarrow}}2\frac{\sigma^{(j)}_z-1}2\frac{\sigma^{(k)}_z-1}2\right.\nonumber\\
&\left.-g_{\uparrow\downarrow}\frac{\sigma^{(j)}_z+1}2\frac{\sigma^{(k)}_z-1}2\right],
\end{align}
where $g_{\sigma\sigma'}=(4\pi a_{\sigma\sigma'}/m)\int d\mathbf r|\phi_0(\mathbf r)|^4$ and $a_{\sigma\sigma'}$ is the s-wave scattering length \cite{Pethick, Bloch}, and $\boldsymbol \sigma^{(j)}$ are the Pauli matrices for the $j$th boson. The interaction Hamiltonian is $\mathcal{PT}$-symmetric only if $g_{\uparrow\uparrow}=g_{\downarrow\downarrow}$; in atomic gases, this condition can be experimentally fulfilled by the technique of Feshbach resonance \cite{Chin}. We focus on this situation $g\equiv g_{\uparrow\uparrow}=g_{\downarrow\downarrow}$ in the following discussion.

Therefore, the total Hamiltonian of $N$ interacting bosons is given by 
\begin{align}
H=&H_{\mathcal PT}+C\label{h}\\
H_{\mathcal PT}=&-2i\Gamma S_z+2JS_x+\delta g\left(S_z\right)^2\label{hpt}\\
C=&-i N\Gamma+\frac g 2 N(N-1)-\frac{\delta g}4 N^2,
\end{align}
where $\boldsymbol S=\sum_{j=1}^N\boldsymbol\sigma^{(j)}/2$, 
and $\delta g\equiv g-g_{\uparrow\downarrow}$. 
Compared with the noninteracting case, the $\mathcal{PT}$-symmetry breaking transition of the interacting boson system is now determined by $H_{\mathcal PT}$.
Since there are two internal states $\ket\uparrow$ and $\ket\downarrow$ to accommodate $N$ bosons, the dimension of the Hilbert space shall be $N+1$. It is easy to assure oneself that such a space is spanned by the streched states $|N/2, m\rangle$ with $m=-N/2,-N/2+1,\dots,N/2-1,N/2$ where $S^2|N/2, m\rangle=N(N+2)/4|N/2, m\rangle$ and $S_z|N/2, m\rangle=m|N/2, m\rangle$. In this space, the matrix element of $H_{\mathcal PT}$ becomes
\begin{align}
\left(H_{\mathcal PT}\right)_{mm'}\equiv &\langle N/2, m|H_{\mathcal PT}|N/2,m'\rangle\nonumber\\
=&(-2i\Gamma \,m+\delta g \, m^2)\delta_{m,m'}\nonumber\\
&+J\sqrt{(N/2)(N/2+1)-m^2+m}\,\delta_{m,m'+1}\nonumber\\
&+J\sqrt{(N/2)(N/2+1)-m^2-m}\,\delta_{m,m'-1}.
\end{align}
The transition occurs when some eigenvalues of $H_{\mathcal PT}$ coalesce and turn complex afterwards. Note that when $\delta g=0$, the interactions drop out of $H_{\mathcal PT}$; this is the situation that the scattering lengths $a_{\sigma\sigma'}$ become all the same and the interatomic interactions are SU(2) invariant. This dependence of the transition on the interactions is because if the interatomic interactions are SU(2) invariant, i.e., $[\mathbf S, H_{int}]=0$, the noninteracting $\mathcal{PT}$-symmetric non-Hermitian Hamiltonian $-2i\Gamma S_z+2JS_x$ and $H_{int}$ are commutable and can be diagonalised simultaneously.

The above non-Hermitian Hamiltonian formalism is related to the Lindbald equation describing the bosons subject to pure loss in the following way. In terms of the field operator $b_\sigma$ ($b^\dagger_\sigma$) which annihilates (creates) a boson of internal state $\sigma$ in the ground state of the harmonic trap, the Lindbald equation for the density matrix $\rho$ of the bosons is given by \cite{Cohen}
\begin{align}
\frac{d\rho}{d t}=-i[\mathcal H_s,\rho]-2\Gamma(b^\dagger_\uparrow b_\uparrow\rho+\rho b^\dagger_\uparrow b_\uparrow)+4\Gamma b_\uparrow\rho b^\dagger_\uparrow,\label{lindbald}
\end{align} 
where $\mathcal H_s$ is the Hermitian Hamiltonian of the bosons in the absence of the external lossy laser coupling. Given that in experiment the initial density matrix $\rho(0)$ shall be always block diagonalized in the number of bosons, i.e., $\rho_{\alpha\beta}(t=0)$ is zero unless $\alpha=\beta$, where $\rho_{\alpha\beta}(t)\equiv P_{\alpha}\rho(t) P_{\beta}$ and $P_\alpha$ is the $\alpha$ boson subspace projection operator, so is $\rho(t)$. 
If the last term $4\Gamma b_\uparrow\rho b^\dagger_\uparrow$ in Eq.~(\ref{lindbald}) were not there, the time dependent density matrix would be given by $\rho(t)=U(t)\rho(0)U^\dagger(t)$ with $U(t)=e^{-i\mathcal H t}$ and $\mathcal H\equiv\mathcal H_s-i 2\Gamma b^\dagger_\uparrow b_\uparrow$; since the projection of $\mathcal H$ in the $N$ boson subspace is just the non-Hermitian Hamiltonian $H$ in Eq.~(\ref{h}), i.e., $H=P_N \mathcal H P_N$, the properties of $H$ would determine the time evolution of $\rho(t)$.  

To access the importance of the term $4\Gamma b_\uparrow\rho b^\dagger_\uparrow$ in Eq.~(\ref{lindbald}) to a typical experiment starting with $N$ particles, we note that initially only $\rho_{NN}(t=0)$ is nonzero.  Since the pure loss can only cause the particle number to decrease, for all the following time, $\rho_{\alpha\beta}(t)=0$ if $\alpha>N$ or $\beta>N$. Thus, from Eq.~(\ref{lindbald}) one can first obtain $\rho_{NN}(t)=U(t)\rho_{NN}(0)U^\dagger(t)$; the term $4\Gamma b_\uparrow\rho b^\dagger_\uparrow$ has no effects on $\rho_{NN}(t)$ since the projection of the term involves only $\rho_{N+1,N+1}(t)$ which is identically zero. Note that $\mathcal H$ commutes with the total particle number. From hereon, one can show $\rho_{N-1,N-1}(t)=\int_0^t d\tau  U(t-\tau) [4\Gamma b_\uparrow\rho_{NN}(\tau) b^\dagger_\uparrow]
U^\dagger(t-\tau)$. Likewise, one can solve all the rest $\rho_{\alpha\alpha}(t)$ for $\alpha<N$ in a cascade; the non-diagonal parts are always zero, i.e., $\rho_{\alpha\beta}=0$ for $\alpha\neq\beta$. The above argument justifies one to study the time evolution of the purely lossy system by analyzing the non-Hermitian Hamiltonian $H$ from Eq.~(\ref{h}). Of course, calculations of observables should resort to the density matrix $\rho(t)$. This justification shall also apply to other similar purely lossy systems. 

\begin{figure}[t]
			\includegraphics[width=3.5 in]{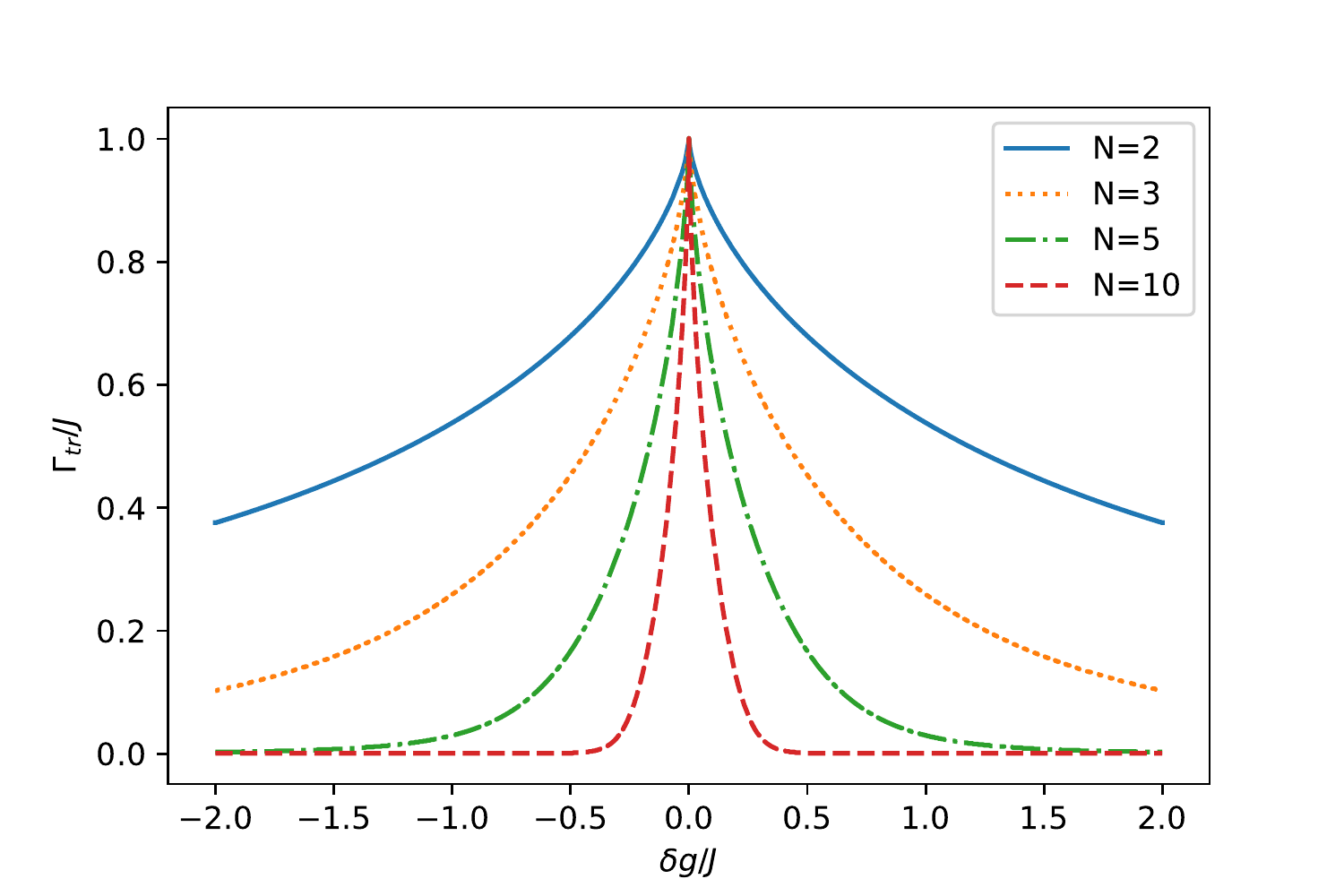}
			\caption{The critical value of $\Gamma_{\rm tr}$ for the $\mathcal{PT}$-symmetry transition versus the interaction parameter $\delta g$. As the magnitude of $\delta g/J$ increases, $\Gamma_{\rm tr}/J$ is suppressed from unity; $\Gamma_{\rm tr}/J$ is symmetric in $\delta g/J$. For fixed $\delta g/J$, $\Gamma_{\rm tr}/J$ acquires a smaller value for larger boson number $N$. 			}
			\label{gamma_tr}
		\end{figure}	
\begin{figure}[t]
			\includegraphics[width=3.5 in]{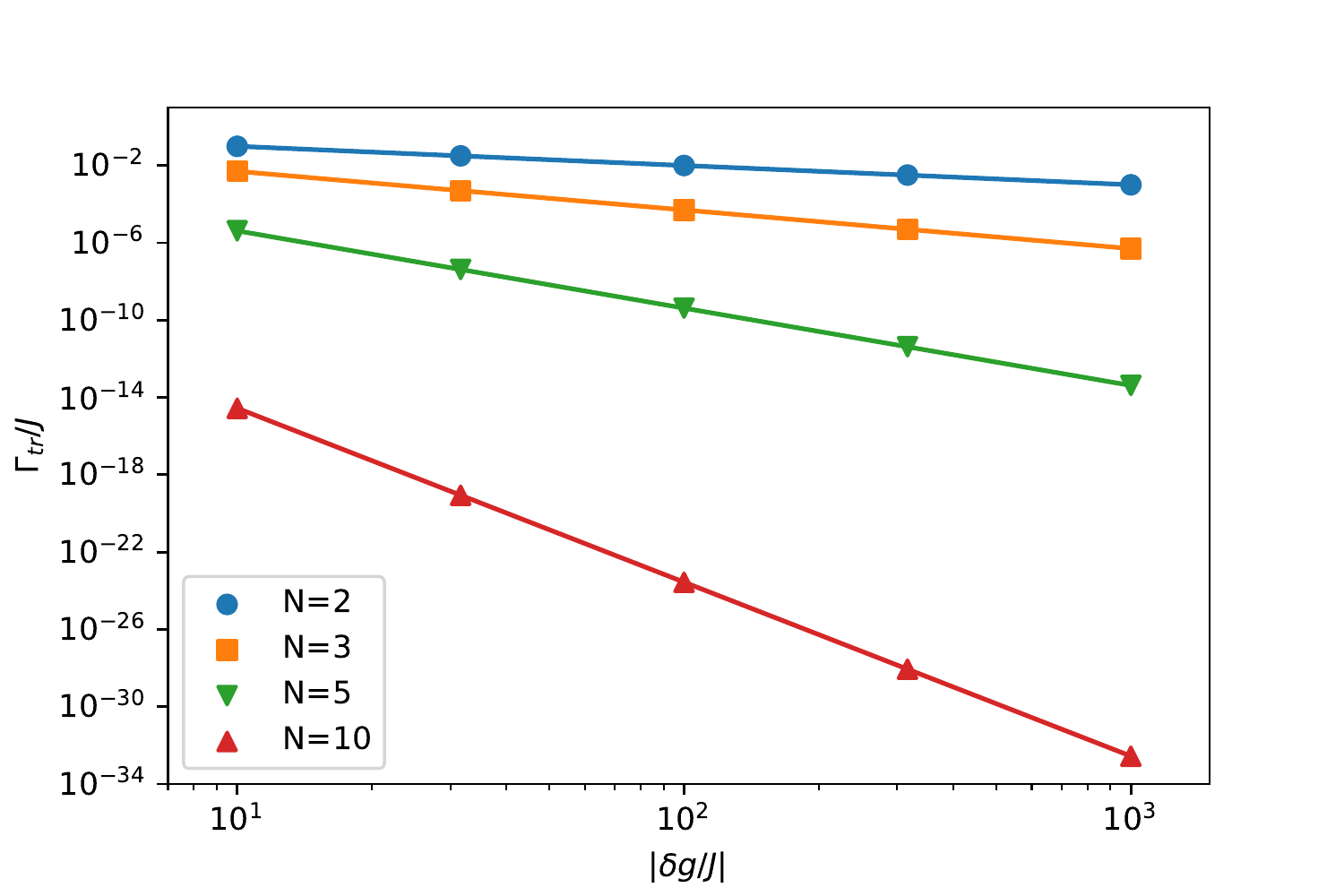}
			\caption{The asymptotic behavior of $\Gamma_{\rm tr}$ in the large $|\delta g|$ limit. The fit to the numerically calculated points gives $\Gamma_{\rm tr}/J\sim|\delta g/J|^{-(N-1)}$, agreeing with the argument given in the text.}
			\label{asymp}
		\end{figure}

We start with  analyzing the non-Hermitian Hamiltonian $H$ of two interacting bosons. For $N=2$, the Hamiltonian $H_{\mathcal PT}$ in the basis $\{\ket\uparrow\ket\uparrow, (\ket\uparrow\ket\downarrow+\ket\downarrow\ket\uparrow)/\sqrt2, \ket\downarrow\ket\downarrow\}$ has the explicit form
\begin{align}
H_{\mathcal PT}=\left[\begin{matrix}-2i\Gamma+\delta g & \sqrt2 J & 0\\
\sqrt2 J & 0 & \sqrt2 J\\
0 & \sqrt2 J & 2i\Gamma+\delta g
\end{matrix}\right].\label{h2}
\end{align}
The corresponding characteristic polynomial is 
\begin{align}
f(\lambda)&\equiv\det[H_{\mathcal PT}-\lambda]\nonumber\\
&=-\lambda^3+2\delta g \lambda^2+[4(J^2-\Gamma^2)-\delta g^2]\lambda-4J^2\delta g,
\end{align}
whose zeros are the eigenvalues of $H_{\mathcal PT}$.
Since all the coefficients of the cubic polynomial $f(\lambda)$ are real, one of the three zeros of $f(\lambda)$ is real definite. When $\Gamma$ is increased from zero, the rest two zeros of $f(\lambda)$, which are also real in the first place, coalesce at the the $\mathcal{PT}$-symmetry transition point and become complex afterwards. This coalescence occurs when the discriminant of $f(\lambda)$
\begin{align}
\Delta(J,\Gamma,\delta g)=&16\big[16(J^2-\Gamma^2)^3+(J^4-20J^2\Gamma^2-8\Gamma^4)\delta g^2\nonumber\\
&-\Gamma^2\delta g^4\big]\label{delta}
\end{align}
is zero. When $\delta g=0$, $\Delta(J,\Gamma_{\rm tr},0)=0$ retrieves the known transition point $\Gamma_{\rm tr}/J=1$. 
For nonzero $\delta g$, Fig.~(\ref{gamma_tr}) shows that $\Gamma_{\rm tr}/J$ is suppressed more and more as $|\delta g/J|$ increases. By Eq.~(\ref{delta}), it is manifest from that $\Gamma_{\rm tr}/J$ is even in $\delta g/J$, and one finds $(\Gamma_{\rm tr}/J)^2\approx 1-3 |\delta g/J|^{2/3}/2^{4/3}$ for $|\delta g/J|\ll1$, and $\Gamma_{\rm tr}/J\approx  |\delta g/J|^{-1}$ for $|\delta g/J|\gg1$. 

The suppression of $\Gamma_{\rm tr}/J$ in the limit $|\delta g/J|\gg1$ is readily understood by inspecting the $\mathcal{PT}$-symmetric Hamiltonian, Eq.~(\ref{h2}). In such a limit, we recast $H_{\mathcal PT}=H_{\mathcal PT,L}+V$ with
\begin{align}
H_{\mathcal PT,L}=&\left[\begin{matrix}\delta g & 0 & 0\\
0 & 0 & 0\\
0 & 0 & \delta g
\end{matrix}\right],\nonumber\\
V=&\left[\begin{matrix}-2i\Gamma & \sqrt2 J & 0\\
\sqrt2 J & 0 & \sqrt2 J\\
0 & \sqrt2 J & 2i\Gamma
\end{matrix}\right]\label{h2sum}.
\end{align}
To the the leading order, $H_{\mathcal PT,L}$ yields right away that the two states $\ket\uparrow\ket\uparrow$ and $\ket\downarrow\ket\downarrow$ are degenerate and share the same eigenvalue $\delta g$, and the eigenvalue of the third state  $(\ket\uparrow\ket\downarrow+\ket\downarrow\ket\uparrow)/\sqrt2$ is zero, well separated apart from $\delta g$. Since the $\mathcal{PT}$-symmetry transition is expected to happen at the point where eigenvalues of $H_{\mathcal PT}$ coalesce, to find the effects of $\Gamma$ and $J$ on the two degenerate eigenvalues originally equal to $\delta g$, we use $V$ to carry out a perturbation calculation to derive the effective Hamiltonian in the subspace spanned by the states $\ket\uparrow\ket\uparrow$ and $\ket\downarrow\ket\downarrow$; we find that to second order of $V$ the effective Hamiltonian is given by \cite{Cohen}
\begin{align}
H_{eff}=&\delta g+2J^2/\delta g+\left[\begin{matrix}  -2i\Gamma & 2J^2/\delta g \\
2J^2/\delta g & 2i\Gamma\end{matrix}\right].\label{heff}
\end{align}
This effective Hamiltonian yields $\Gamma_{\rm tr}/J\approx|J/\delta g|^{-1}$, the same as from requiring $\Delta(J,\Gamma_{\rm tr},\delta g)=0$.

For $N>2$, we numerically diagonalise $H_{\mathcal PT}$ and find that the $\mathcal{PT}$ transition is always due to the coalescence of a pair of eigenvalues of $H_{\mathcal PT}$. Figure~(\ref{gamma_tr}) shows that the critical value $\Gamma_{\rm tr}/J$ is also symmetric in $\delta g/J$, which is because, under the transformation $\delta g\to -\delta g$ and $\mathbf S\to-\mathbf S$, we have $H_{\mathcal PT}\to -H_{\mathcal PT}$.
We find that for fixed $N$, $\Gamma_{\rm tr}/J$ decreases as $|\delta g/J|$ increases, while for fixed $\delta g/J$, as $N$ increases, $\Gamma_{\rm tr}/J$ is more and more suppressed. 

Figure (\ref{asymp}) shows that in the large $|\delta g/J|$ limit, $\Gamma_{\rm tr}/J\sim|\delta g/J|^{-(N-1)}$. This asymptotic behavior can be understood by an analysis similar to the one given above for $N=2$.  For arbitrary $N$, in the large $|\delta g|$ limit, we separate $H_{\mathcal PT}$ from Eq.~(\ref{hpt}) as $H_{\mathcal PT}=H_{\mathcal PT, L}+V$ with $H_{\mathcal PT, L}=
\delta g\left(S_z\right)^2$ and $V=-2i\Gamma S_z+2JS_x$. In such a limit, the leading order Hamiltonian $H_{\mathcal PT, L}$ gives rise to a pair of degenerate eigvenvalues in each subspaces spanned by $|N/2, m\rangle$ and $|N/2, -m\rangle$; the eigenvalues $m^2\delta g$ are all well separated from each other. To determine the transition, we use $V$ to derive the effective Hamiltonian $H_{eff}$ in the each two dimensional subspace. It is easy to convince oneself that to the lowest order of $J$, the diagonal elements are $\langle N/2, \pm m|H_{eff}|N/2, \pm m\rangle=m^2\delta g\mp2im \Gamma$, and the off-diagonal elements are generated at order of $V^{2|m|}$, resulting in $\langle N/2, \pm m|H_{eff}|N/2, \mp m\rangle\sim J^{2|m|}/\delta g^{2|m|-1}$. Thus, by diagonalizing the $2\times2$ matrix of $H_{eff}$, we find the transition point $\Gamma_{\rm tr}/J\sim|\delta g/J|^{-(2|m|-1)}$ for each two dimensional subspaces. Given that the maximum value of $|m|$ equals $N/2$, overall, the $N$-body system enters into the symmetry breaking phase first at $\Gamma_{\rm tr}/J\sim|\delta g/J|^{-(N-1)}$.

\begin{figure}[t]
			\includegraphics[width=3.5 in]{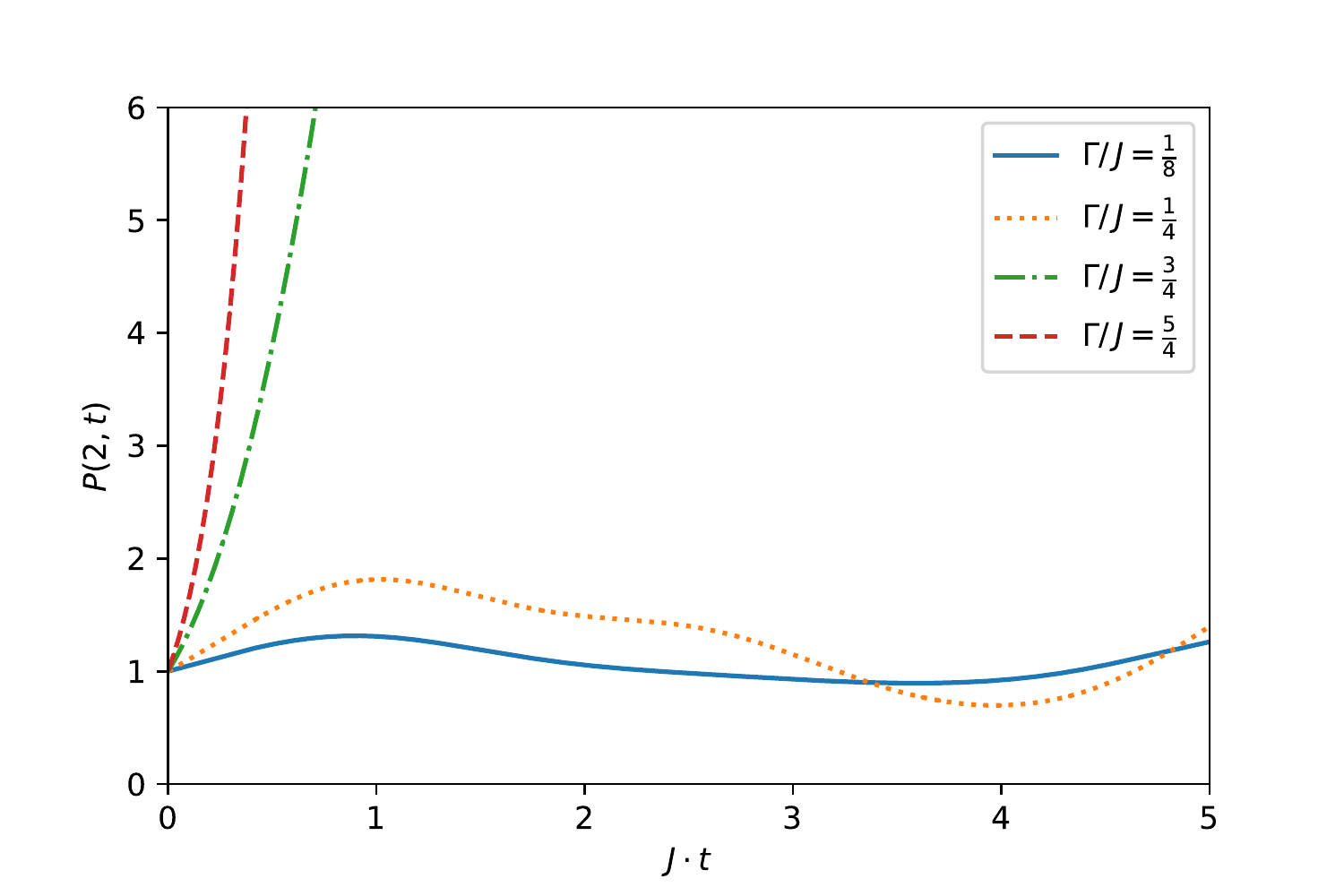}
			\caption{The rescaled probability ${\rm P}(2,t)$ of finding two bosons in an experiment with initially two bosons and $\delta g/J=1$ for various values of $\Gamma/J$. The cases of $\Gamma/J=1/8,1/4$ are in the $\mathcal{PT}$-symmetric phase, and ${\rm P}(2,t)$ is bounded. The cases of $\Gamma/J=3/4,5/4$ are in the $\mathcal{PT}$-symmetry breaking phase, and ${\rm P}(2,t)$ grows exponentially.}
			\label{p2}
		\end{figure}

The relation between the non-Hermitian Hamiltonian formalism and the Lindbald equation for our system given above indicates that the signatures of the $\mathcal{PT}$-symmetric and symmetry breaking phases governed by $H_{\mathcal PT}$ in Eq.~(\ref{hpt}) can be detected experimentally in the following way. Let one prepare the experiment initally with $N$ bosons \cite{Bloch2010,Greiner2010,Bloch2011} such that $\rho_{NN}(t)=e^{-2N\Gamma t} e^{-iH_{\mathcal PT}t}\rho_{NN}(0)e^{iH_{\mathcal PT}^\dagger t}$; the quantity $e^{2N\Gamma t} \rho_{NN}(t)$ shall have qualitatively different time dependent behaviors in the symmetric and symmetry breaking phases. For example, by the high accuracy atom number detection achieved experimentally \cite{Jochim}, one can measure the rescaled probability of finding $N$ bosons ${\rm P}(N,t)=e^{2N\Gamma t}\tilde {\rm P}(N,t)$ with $\tilde{\rm P}(N,t)\equiv {\rm Tr}\rho_{NN}(t)$. In contrast, the total number of atoms was measured to distinguish the two phases for the noninteracting $^6$Li atoms \cite{Luo}. In our interacting case, the total number of atoms ceases to be a good observable for the purpose since the observable depends on not only  $\rho_{NN}(t)$ but also $\rho_{\alpha\alpha}(t)$ for $\alpha<N$ whose dynamics is not determined by a single Hamiltonian $H$ for $\alpha$ bosons.
Figure (\ref{p2}) plots the rescaled ${\rm P}(2,t)$ in an experiment starting with $N=2$ bosons and $\delta g/J=1$ for various values of $\Gamma/J$. Note that for $N=2$ and $\delta g/J=1$, $\Gamma_{\rm tr}/J\approx0.538$. Figure (\ref{p2}) shows that ${\rm P}(2,t)$ is bounded in the $\mathcal{PT}$-symmetric phase, and grows exponentially in the symmetry breaking one. Due to the interactions, the point $\Gamma/J=3/4$ is already in the symmetry breaking phase while its value is still smaller than the critical value $\Gamma_{\rm tr}/J=1$ for the noninteracting case.

\emph{Acknowledgements.} We thank Jiaming Li for discussions. This work is supported by NSFC Grants No. 11474179, No.
11722438, and No. 91736103.

\end{document}